\documentclass[aps,prb,twocolumn,showpacs,superscriptaddress]{revtex4-1}

\usepackage{times}
\usepackage{color}
\usepackage{graphicx}% Include figure files
\usepackage{dcolumn}% Align table columns on decimal point
\usepackage{amssymb}
\usepackage{amsmath}
\usepackage{amsfonts}
\usepackage{docs}%
\usepackage{bm}%
\usepackage{wrapfig}
 \usepackage{multirow}

\usepackage[colorlinks=true,linkcolor=blue,pagecolor=blue,filecolor=blue,menucolor=blue,urlcolor=blue,citecolor=blue,anchorcolor=blue]{hyperref}%

\newcommand{\sns}{$S$/$N$/$S$ }
\newcommand{\sfs}{$S$/$F$/$S$ }

\newcommand{\fs}{$S$/$F$ }

\newcommand{\f}{$F$ }
\newcommand{\s}{$S$ }
\newcommand{\ns}{$N$/$S$ }

\begin{document}

\title{Spontaneous Edge Accumulation of Spin Currents in Finite-Size Two-Dimensional Diffusive Spin-Orbit Coupled $SFS$ Heterostructures}

\author{Mohammad Alidoust }
\email{phymalidoust@gmail.com} 
\affiliation{Department of Physics,
University of Basel, Klingelbergstrasse 82, CH-4056 Basel, Switzerland}
\affiliation{Department of Physics,
Faculty of Sciences, University of Isfahan, Hezar Jerib Avenue,
Isfahan 81746-73441, Iran}
\author{Klaus Halterman}
\email{klaus.halterman@navy.mil} \affiliation{Michelson
Lab, Physics Division, Naval Air Warfare Center, China Lake,
California 93555, USA}

\date{\today}

\begin{abstract} 
We theoretically study spin and charge currents through
finite-size two-dimensional $s$-wave superconductor/uniform ferromagnet/$s$-wave
superconductor ($S$/$F$/$S$) junctions with intrinsic spin-orbit
interactions (ISOIs) using a quasiclassical approach. Considering
experimentally realistic parameters, we demonstrate that the combination of
spontaneously broken time-reversal symmetry and lack of inversion symmetry can result
in spontaneously accumulated spin currents at the edges of finite-size
two-dimensional magnetic \fs
hybrids.
Due to the spontaneous edge spin accumulation, the corners of the $F$ wire
host the maximum spin current density.
We further
reveal that
this type edge phenomena are robust and independent of either the actual type of
ISOIs or  exchange field orientation. Moreover, we study spin
current-phase relations in these diffusive spin-orbit coupled \sfs junctions.
Our results unveil net spin currents, not
accompanied by charge supercurrent, that spontaneously accumulate at
the sample
edges through a modulating superconducting phase difference. Finally,
we discuss possible experimental implementations to observe these edge phenomena.
\end{abstract}


\maketitle

\section{Introduction}\label{sec:intro}
Spintronics devices operate by spin transport
mechanisms \cite{prinz_1,wolf_1,zhang_1,shen_1,sharma_1,kikkawa_1}
rather than by
utilizing charged carriers, as is done typically in
conventional electronics devices.
The use of spin currents   can result in higher speeds and reduced
dissipation\cite{sharma_1,kikkawa_1} while
exhibiting weak sensitivity to nonmagnetic impurities and temperature.\cite{sharma_1,shen_1,zhang_1}
For functional spin-based devices, it is necessary to manipulate and generate
spin-currents in a practical and efficient manner. For this reason, many investigations have focused on harnessing the
spin-orbit interactions\cite{rashba_term,dresselhaus_term} (SOIs) present in many materials, including
semiconductors.\cite{major1,major2,major3,major4,major5,Wunderlich_1,popovic,Miron_1,Garello_1}
The SOI is a
quantum relativistic phenomenon that can be divided into two
categories: $i)$ intrinsic (originating from the electronic band
structure of the material) and $ii)$ extrinsic (originating from
spin-dependent scattering of
impurities).\cite{sharma_1,shen_1,zhang_1}
The intrinsic spin-orbit interactions (ISOIs) such
as Rashba \cite{rashba_term} and Dresselhaus
\cite{dresselhaus_term}, are experimentally controllable via tuning
a gate voltage
\cite{rashba_book_1,grundler_1,nitta_1,winklwer_1,awschalom_1,Miron_1,Erlingsson,tanaka_2}.
This particular attribute  has proliferated efforts
striving for
high-performance spin-based devices, including  transistors, and
new routes in information storage and transport.\cite{wolf_1,stepanenko_1,dery_1,rashba_book_1,Lee_1,Wunderlich_1,Miron_1}
Similarly, ferromagnet ($F$) and superconductor ($S$) heterostructures
have received  renewed interest lately due to the possibility of
generating spin polarized triplet supercurrents\cite{bergeret1,buzdin_rmp,Kontos,
robinson,DOS_2_ex,alidoust_mgr,
Fominov,Sosnin,spnactv_21,halterman1} that can be used for practical purposes\cite{alidoust_mgr}.
By considering a ferromagnet with an ISOI,
the spin orbit interaction can couple with the magnetic exchange field, resulting in
modified superconducting proximity effects and
additional venues for
new spin phenomena.
Indeed, the ISOI can induce
long-range proximity effects  in uniformly
magnetized \fs structures
due to the momentum-dependence of the
effective exchange field\cite{bergeret_so}.
It is therefore of fundamental importance not only
to find a simple, experimentally accessible structure that
can support spin currents in $F/S$ systems,
but it is also crucial to
determine the
spatial behavior of the spin currents
near the boundaries of the superconducting hybrids.

Many past works are based on
the application of {\it external} electric or magnetic fields.
One of the earliest such cases involved
the combination of SOIs and an  external electric field,
giving rise to an accumulation of spin
currents at the edges\cite{chazalviel,hirsh_1}
 (the so called spin-Hall effect\cite{Sinova_1,Murakami_1}).
The spin
currents generally tend to peak near the sample boundaries and vanish at the
electrode/sample interfaces.\cite{mishchenko_1,Nikolic,Onoda,gorini_1}
These theoretical predictions were later experimentally observed
in semiconductor samples\cite{kato_1}.
The spin-Hall phenomenon
was also
extensively studied in superconducting heterostructures
where various types of spin-orbit coupling (SOC) play key roles.
\cite{malsh_sns,malsh_severin,kontani_1,malsh_5,cheng_1,tao_1,franz_1,service_1,reynoso_1,liu_2,sns_so_bezuglyi_1,sns_so_Krive_1,sns_so_dellanna_1,sns_so_beri_1,nazar_1,tanaka_2,bobkova_1,eschrig_1,arahata_1}
For example, the out-of-plane
component of the spin density was theoretically investigated\cite{malsh_severin}
 in a spin-orbit coupled
$S/N/S$
junction [with normal metal ($N$) interlayer] subject to an inhomogeneous external magnetic
field. It was found that the spin density varies along the
transverse
direction, leading to a longitudinal phase difference between the
\s electrodes.
The
influence of extrinsic SOIs on the critical supercurrent in
diffusive magnetic hybrid structures was also studied.\cite{buzdin_so_ext,demler_1}
In nonmagnetic \sns Josephson
junctions with SOC subject to a magnetic field,
 $0$-$\pi$
transitions may be induced.\cite{buzdin_rmp}
In an earlier work\cite{niu_1},
singlet-triplet pair conversion was numerically investigated using a lattice
model in a ballistic half metal ferromagnetic Josephson junction
with an interfacial Rashba SOC.
Several optimal configurations have also been
theoretically studied  for generating and detecting
the predicted long-range triplet correlations in
experiments.\cite{bergeret_so}

In this paper, we study
the local spin currents in uniformly magnetized \sfs Josephson junctions with
spatially 
uniform intrinsic SOIs,
avoiding any external electric or magnetic fields. 
We
employ a two-dimensional quasiclassical
Keldysh-Usadel approach that incorporates a generic spin-dependent
vector potential to study the behavior of the spin current components.
We consider two types of ISOCs: $i)$ Rashba and
$ii)$ Dresselhaus SOC, and the magnetization of the \f wire can
take arbitrary orientations.
We find that the coupling of the \f wire's exchange field and ISOIs leads to
edge spin currents with three nonzero components. The
spin current density components peak near the edges of the \f strip and sharply
decline when moving away from the edges. Therefore, the maximal spin current
accumulation takes place near the \f wire's corners. This phenomenon can
be also observed in ISO coupled \sns junctions with a single spin
active \ns interface as demonstrated in Ref. \onlinecite{ma_kh_soc_sa}. 
Our results show that the spin current can be
switched on or off
at the \fs contacts, depending on the magnetization
direction. The spatially averaged spin current components reveal a
$2\pi$-periodicity and even-functionality in $\varphi$, the phase difference
between the \s terminals.
This is in
contrast to the charge supercurrent which is a $2\pi$-periodic odd function of $\varphi$
(and consistent with previous studies\cite{alidoust_1}).
Note that for such junctions, the
argument in the current phase relation
for some situations can become modified by a simple
$\varphi_0$ shift. \cite{buzdin_phi0,Konschelle,nazar_1}

The simple hybrid structure proposed here relies solely on the intrinsic properties of
the system, in contrast  to
 other structures that rely inextricably on external fields to
observe the edge spin
currents\cite{Murakami_1,hirsh_1,kato_1,Sinova_1,mishchenko_1,chazalviel,
Nikolic,Onoda,malsh_1,malsh_sns,malsh_severin}.
Our device consists of a finite-size intrinsic
SO coupled \f wire (with uniform magnetization) sandwiched between two \s banks.
The spin
currents then spontaneously accumulate  at the
sample edges, \textit{without}
the application of an external electric or magnetic field to the system.
We demonstrate that the device allows for the realization of spin currents in
the \textit{absence} of charge supercurrent
by simply modulating $\varphi$.
The edge spin accumulation is a
signature of the spin Hall effect\cite{malsh_sns}, and hence can be experimentally measured
by optical experiments for instance\cite{kato_1} (see the discussions in Sec.~\ref{sec:results}).
Also, we discuss the symmetries
present among the spin current components when varying the
magnetization orientation with Rashba or Dresselhaus
SOC present. Moreover, we
compare our results with
the
charge and
spin currents found in a nonmagnetic diffusive \sns Josephson junction
with Rashba and/or Dresselhaus SOC. We find that the spin
currents vanish in the \sns devices, consistent with previous
works\cite{malsh_sns,ma_kh_soc_sa}, and that the charge
current displays a spatially uniform profile without any transverse
component, indicating conservation of charge current. 

The paper is organized as follows. We outline the theoretical
framework used to study hybrid structures with ISOCs in Sec.~\ref{sec:theor}.
In Sec.~\ref{sec:results}, the results of diffusive \sfs
Josephson junctions are presented
for the case of Rashba ISOC.
We next utilize the symmetries in the spin currents to give a simple prescription
for finding the corresponding results for the Dresselhaus spin-orbit interaction.
We finally present
concluding remarks in Sec.~\ref{sec:conclusion}.

\section{Theoretical formalism}\label{sec:theor}
The intrinsic SOI is a consequence of the moving carriers' spin
interaction with an atomic potential $V({\bm r})$. Therefore, the
total Hamiltonian of a moving electron in such an atomic potential can
be expressed as,\cite{wu_1,rashba_book_1}
\begin{equation}\label{eq:hamiltonian}
      \mathit{\mathbb{H}}=\frac{\vec{\mathcal{P}}^2}
      {2m}+\frac{e\hbar^2}{4m_0^2c^2}\vec{\mathcal{P}}
      \cdot\Big\{\vec{\tau}\times\vec{\partial}V({\bm
    r})\Big\},
\end{equation}
where $m_0=0.51$Mev,
is the free electron
mass and
 and $c$
 is the velocity of light in vacuum. We define $\vec{\mathcal{P}}$
to represent the electron's momentum vector, and
$\vec{\partial}\equiv(\partial_x,\partial_y,\partial_z)$.
The vector of Pauli matrices,
$\vec{\tau}$, is given in
Appendix \ref{app:pauli}. It has been shown that the
linearized SOC term can be simply accounted for
as an effective background
field that follows SU(2) gauge
symmetries.\cite{bergeret_so,gorini_1,gorini_2,gorini_3}
Hence, it is sufficient to replace partial derivatives, appearing in
the quasiclassical formalism, by their
covariants.\cite{bergeret_so,gorini_1} Another advantage
of the SU(2) approach is the convenient definition of physical
quantities such as spin currents.\cite{duckheim_1}

We start with the Usadel equations that enable us to study the
charge and spin transport through diffusive $F/S$ systems
with the ferromagnetic regions having
arbitrary magnetization patterns $\vec{h}({\bm
r})=\big(h^x({\bm r}),h^y({\bm r}),h^z({\bm r})\big)$:
\cite{bergeret1,Usadel,bergeret_so}
\begin{gather}\label{eq:usadel}
 \Big[\hat{\partial},\hat{G}({\bm r})[\hat{\partial},\hat{G}({\bm
r})]\Big]=\frac{-i}{D}\Big[ \varepsilon \hat{\rho}_{3}+
\text{diag}[\mathcal{H}({\bm r}),\mathcal{H}^{\mathcal{T}}({\bm
r})],\hat{G}({\bm r})\Big],\\\nonumber \mathcal{H}({\bm
r})=\vec{h}({\bm r})\cdot\vec{\sigma},
\end{gather}
where $\vec{\hat{\rho}}$ and $\vec{\sigma}$ denote vectors comprised of
$4\times 4$ and $2\times 2$ Pauli matrices (see Appendix
\ref{app:pauli}), and $D$ represents the diffusive constant of the
ferromagnetic medium.
 We have
denoted the quasiparticles' energy by $\varepsilon$ which is
measured from the Fermi surface $\varepsilon_F$. Throughout this
work, we focus on the low proximity limit of the diffusive regime
\cite{bergeret1}. In this limit, the normal and anomalous components
of the Green's function can be approximated by, $\underline{F}^{no}({\bm
r})\simeq \underline{1}$ and $\underline{F}({\bm r})\ll
\underline{1}$, respectively. Thus, the advanced component of total
Green's function, $\hat{G}({\bm r})$, takes the following form:
\begin{align}
\hat{G}^{A}({\bm r},\varepsilon)\approx\begin{pmatrix}
-\underline{1} & -\underline{F}({\bm r},-\varepsilon)\\
\underline{F}^\ast({\bm r},\varepsilon) & \underline{1}\\
\end{pmatrix},
\end{align}
where each entry stands for a 2$\times$2 matrix block. Considering
the Taylor expansion, the advanced component can be given by:
\begin{align}\label{Advanced Gree}
&\nonumber\hat{G}^{A}({\bm r},\varepsilon)=\\&\begin{pmatrix}
-1 & 0 & -f_{\uparrow\uparrow}({\bm r},-\varepsilon) & -f_{-}({\bm r},-\varepsilon) \\
0 & -1  &-f_{+}({\bm r},-\varepsilon)  & -f_{\downarrow\downarrow}({\bm r},-\varepsilon) \\
 f_{\uparrow\uparrow}^{\ast}({\bm r},\varepsilon)  & f_{-}^{\ast}({\bm r},\varepsilon)  & 1  &  0  \\
f_{+}^{\ast}({\bm r},\varepsilon)& f_{\downarrow\downarrow}^{\ast}({\bm r},\varepsilon)  & 0  &  1  \\
\end{pmatrix}.
\end{align}
Here we restrict our calculations to the equilibrium situations
where the Retarded and Keldysh blocks of total Green's function are
obtained by: $\hat{G}^{A}({\bm r})=-\big\{\hat{\rho}_3\hat{G}^R({\bm
r})\hat{\rho}_3\big\}^{\dag}$, and $\hat{G}^{K}({\bm
r})=\tanh(\varepsilon k_BT/2)\big\{\hat{G}^{R}({\bm
r})-\hat{G}^{A}({\bm r})\big\}$. Here,
$k_B$ and $T$ denote the Boltzmann
constant and system temperature, respectively.

\begin{figure}[t!]
\includegraphics[width=8.0cm,height=3.0cm]{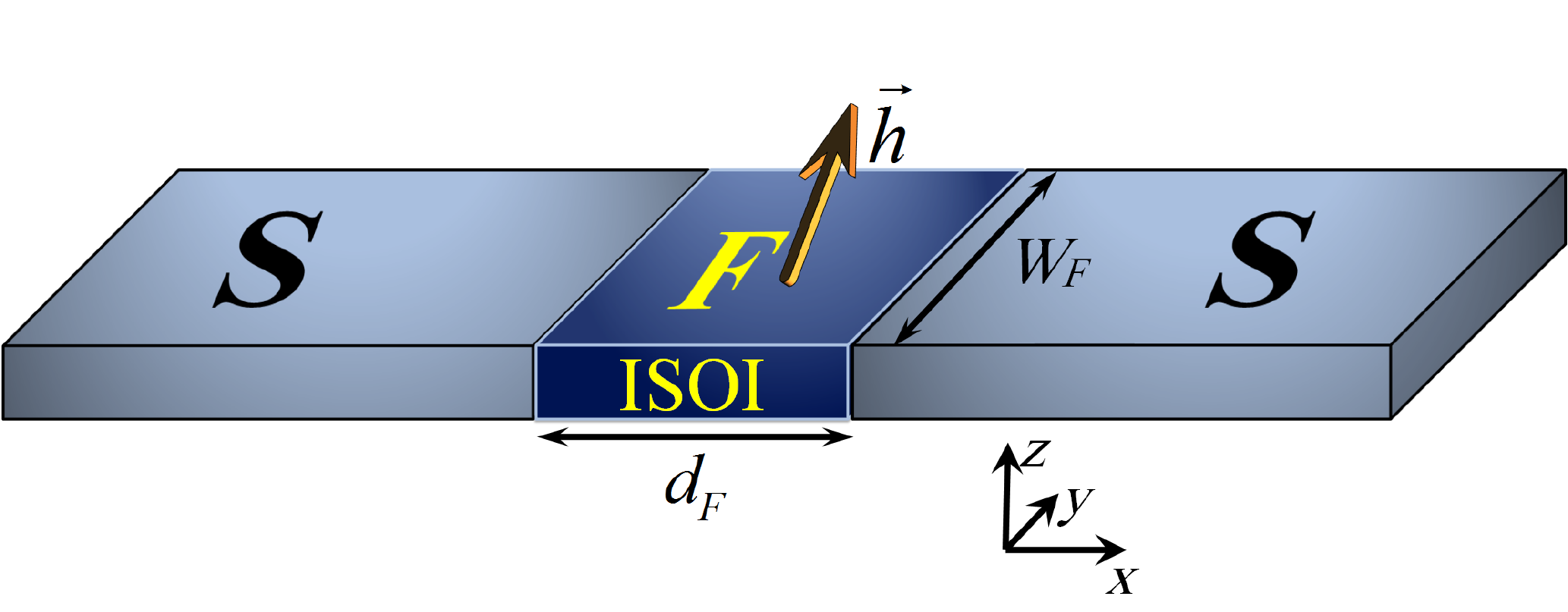}
\caption{\label{fig:model1} (Color online) Schematic of a
finite-size two-dimensional magnetic \sfs Josephson junction.
The
superconducting electrodes and rectangular ferromagnetic nano-wire
are labelled $S$ and $F$, respectively. We assume that the
quasiparticle current experiences an
intrinsic spin-orbit interaction (ISOI)
solely inside the $F$ region.
The thickness and width of
the ferromagnetic strip are labeled $d_F$ and $W_F$, respectively.
The junction is located in the $xy$ plane and the \fs interfaces
are along the $y$ axis. The $F$ region has a uniform exchange field
denoted by $\vec{h}$ and can take arbitrary orientations
$\big(h^x,h^y,h^z\big)$. }
\end{figure}

The Usadel equation, Eq.~(\ref{eq:usadel}), leads to sixteen coupled
complex partial differential equations in the low proximity limit
that become highly complicated with the presence of intrinsic SOI
terms. Unfortunately, the resultant system of coupled
differential equations can only be simplified and decoupled under
very limiting conditions,\cite{buzdin_rmp,bergeret1}
 leading to analytical results. However, for
the  systems considered in this paper, numerical methods are the
most appropriate, and often the only possible routes to investigate
the relevant transport properties.\cite{bergeret_so}
The differential equations must be supplemented by the
appropriate boundary conditions to properly capture the transport
characteristics of \sfs hybrid structures. We thus employ the
Kupriyanov-Lukichev boundary conditions at the \fs interfaces
\cite{cite:zaitsev} and control the intensity of induced proximity
correlations using the barrier resistance parameter, $\zeta$:
\begin{equation}\label{eq:bc}
    \zeta\big\{\hat{G}({\bm r})\hat{\partial}\hat{G}({\bm r})\big\}\cdot\hat{\boldsymbol{n}}=[\hat{G}_{\text{BCS}}(\theta),\hat{G}({\bm
    r})],
\end{equation}
where $\hat{\boldsymbol{n}}$ is a unit vector, directed
perpendicular to a given interface. The solutions to Eqs.~(\ref{eq:usadel})
for a bulk, even-frequency $s$-wave superconductor reads,
\begin{align}
\hat{G}^{R}_{\text{BCS}}(\theta)=\left(
                                  \begin{array}{cc}
                                    \underline{1}\cosh\vartheta(\varepsilon) & i\sigma_2e^{i\theta}\sinh\vartheta(\varepsilon) \\
                                    i\sigma_2e^{-i\theta}\sinh\vartheta(\varepsilon) & -\underline{1}\cosh\vartheta(\varepsilon) \\
                                  \end{array}
                                \right),
\end{align}
in which,
\begin{equation}
\nonumber
\vartheta(\varepsilon)=\text{arctanh}(\frac{\mid\Delta\mid}{\varepsilon}),
\end{equation}
is defined in terms of the superconducting gap $\Delta$.
Here the macroscopic phase of the bulk superconductor is denoted by
$\theta$, so that the difference between the macroscopic phases of the left and
right \s electrodes are given by $\theta_l-\theta_r=\varphi$. For
more compact expressions in our subsequent calculations, we
define the following piecewise functions:
\begin{eqnarray}
&&\nonumber s(\varepsilon)\equiv e^{i\theta}\sinh\vartheta(\varepsilon)=\\&&\nonumber-\Delta\left\{\frac{\text{sgn}(\varepsilon)}{\sqrt{\varepsilon^2-\Delta^2}}\Theta(\varepsilon^2-\Delta^2)-\frac{i}{\sqrt{\Delta^2-\varepsilon^2}}\Theta(\Delta^2-\varepsilon^2)\right\},\\
&&\nonumber
c(\varepsilon)\equiv\cosh\vartheta(\varepsilon)=\\&&\nonumber\frac{\mid\varepsilon\mid}{\sqrt{\varepsilon^2-\Delta^2}}\Theta(\varepsilon^2-\Delta^2)-\frac{i\varepsilon}{\sqrt{\Delta^2-\varepsilon^2}}\Theta(\Delta^2-\varepsilon^2),
\end{eqnarray}
where $\Theta(x)$ stands for the usual step function.
It is clear that the
general boundary conditions given by Eq.~(\ref{eq:bc})
do not permit
current flow  through the hard wall boundaries of the
finite-size two-dimensional \sfs Josephson junction, shown in Fig.~\ref{fig:model1}.

%------------------------------------------- figure 1 -------------------------
%--------------------------------------------------------------------------

To study the influence of differing types of ISOI on the system
transport characteristics, we adopt a spin-dependent tensor vector
potential $\vec{A}({\bm r})=\big(A_x({\bm r}),A_y({\bm r}),A_z({\bm
r})\big)$,
as
follows:\cite{bergeret_so,gorini_1,gorini_2,gorini_3,ma_kh_soc_sa}
\begin{subequations}\label{eq:full_vec_potential}
\begin{align}
A_x({\bm r})=\frac{1}{2}\Big\{\mathcal{A}_x^x({\bm r})\tau^x+\mathcal{A}_x^y({\bm r})\tau^y+\mathcal{A}_x^z({\bm r})\tau^z\Big\},\\
A_y({\bm r})=\frac{1}{2}\Big\{\mathcal{A}_y^x({\bm r})\tau^x+\mathcal{A}_y^y({\bm r})\tau^y+\mathcal{A}_y^z({\bm r})\tau^z\Big\},\\
A_z({\bm r})=\frac{1}{2}\Big\{\mathcal{A}_z^x({\bm
r})\tau^x+\mathcal{A}_z^y({\bm r})\tau^y+\mathcal{A}_z^z({\bm
r})\tau^z\Big\}.
\end{align}
\end{subequations}
Using the above vector potential, we define the covariant
derivatives by;
\begin{equation}\label{eq:partial_derivativ}
    \hat{\partial}\equiv\vec{\partial} \hat{1}-ie
\vec{A}({\bm r}).
\end{equation}
Accordingly, the brackets seen in the Usadel equation, Eq.~(\ref{eq:usadel}), and the boundary conditions, Eq. ({\ref{eq:bc}}),
(as well as the charge and spin currents that shall be discussed
below, Eqs. (\ref{eq:chargecurrentdensity}) and
(\ref{eq:spincurrentdensity})) take the following form:
\begin{equation}\label{eq:brackets}
[\hat{\partial},\hat{G}({\bm r})]=\vec{\partial}\hat{G}({\bm
r})-ie[\vec{A}({\bm r}),\hat{G}({\bm r})].
\end{equation}

The spin and charge currents are key quantities that lend insight
into the fundamental system transport aspects
that provide valuable and crucial information
for nanoscale elements
 in  superconducting spintronics devices, as
described in the introduction. Under equilibrium
conditions, the vector charge ($\vec{J}^c$) and spin
($\vec{J}^{s\gamma}$) current densities can be expressed by the
Keldysh block as follows:\cite{gorini_1,gorini_3}
\begin{equation}\label{eq:chargecurrentdensity}
\vec{J}^c({\bm r},\varphi) = J_{0}^c
\bigg|\int_{-\infty}^{+\infty}\hspace{-.2cm}
d\varepsilon\text{Tr}\Big\{\hat{\rho}_{3} \big(\check{G}({\bm
r})[\check{\partial},\check{G}({\bm r})]\big)^{K}\Big\}\bigg|,
\end{equation}
\begin{equation}\label{eq:spincurrentdensity}
\vec{J}^{s\gamma}({\bm r},\varphi) = J_{0}^s
\bigg|\int_{-\infty}^{+\infty}\hspace{-.2cm}
d\varepsilon\text{Tr}\Big\{\hat{\rho}_{3}
\nu^{\gamma}\big(\check{G}({\bm r})[\check{\partial},\check{G}({\bm
r})]\big)^{K}\Big\}\bigg|,
\end{equation}
where $J_{0}^c  =  N_{0} e D/4$, $J_{0}^s=\hbar J_{0}^c/2e$, and
$N_{0}$ is the number of states at the Fermi surface. The vector
current densities determine the local direction and amplitude of the
currents as a function of coordinates inside the $F$ strip. In other
words, $\vec{J}({\bm r})$, provides a spatial map to the currents
inside the system. We designate $\gamma=x,y,z$ for the three
components of spin current, $\vec{J}^{s\gamma}$. The matrices
we use throughout our derivations are given in Appendix \ref{app:pauli}.
To obtain the total Josephson charge current flowing through the
magnetic strip, an additional integration over the $y$ direction
should be performed on Eq.~(\ref{eq:chargecurrentdensity}) (see Fig.
\ref{fig:model1}). The spin-dependent fields
yield lengthly and cumbersome expressions,
the details of which are not presented here for clarity.
Having now outlined the theoretical approach utilized in
this paper, we can now present our findings in the next section.

\section{Results and discussions}\label{sec:results}
In our computations below, we consider a uniform and
coordinate-independent vector potential, $\vec{A}({\bm r})$, i.e.
$\vec{\partial}\cdot\vec{A}({\bm r})=0$, so that  the spin
vector potential is constant in the entire \f region. A
specific choice for the constant spin vector potential that results
in Rashba ($\alpha$) \cite{rashba_term} and Dresselhaus ($\beta$)
\cite{dresselhaus_term} types of SOC is,
\begin{align}\label{eq:Ax_Ay_Az}
    \left\{\begin{array}{l}
      \mathcal{A}_x^x=-\mathcal{A}_y^y=2\beta, \\
      %\hline
      \mathcal{A}_x^y=-\mathcal{A}_y^x=2\alpha, \\
      \hline
      \mathcal{A}_x^z=\mathcal{A}_y^z=0,\\
      %\hline
      \mathcal{A}_z^z=\mathcal{A}_z^x=\mathcal{A}_z^y=0.
    \end{array}\right.
\end{align}
By substituting the above set of parameters into Eqs.~(\ref{eq:full_vec_potential}),
we arrive at,
\begin{subequations}\label{eq:rash_dress}
\begin{align}
A_x=\beta\tau^x-\alpha\tau^y,\\
A_y=\alpha\tau^x-\beta\tau^y.
\end{align}
\end{subequations}
%------------------------------------------- figure 2 -------------------------
\begin{figure}[b!]
\includegraphics[width=7.50cm,height=6.20cm]{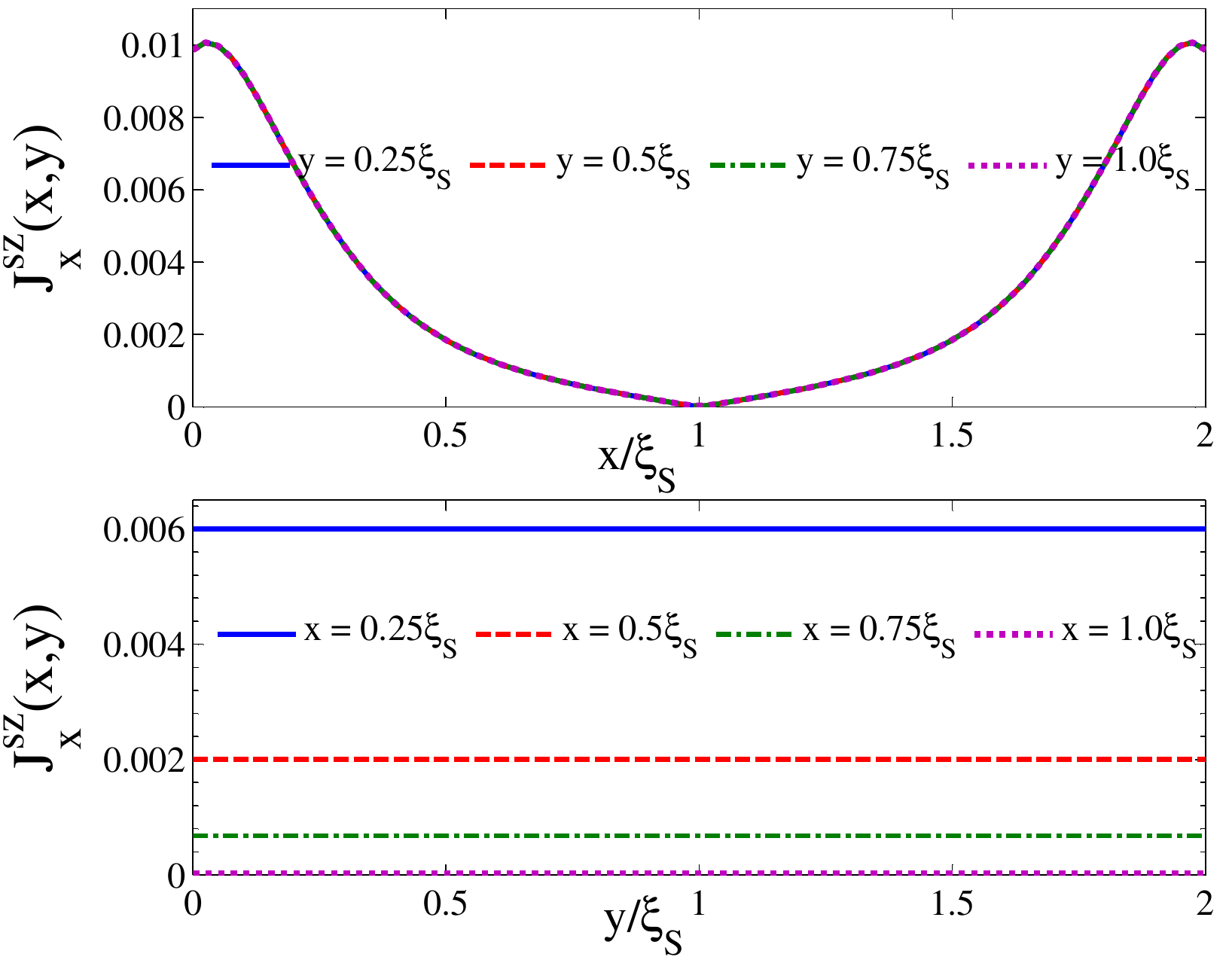}
\caption{\label{fig:hz_10_beta_0_alpha_0_xy} (Color online) Spatial
profile of the spin current in a uniformly magnetized \sfs Josephson (see Fig.~\ref{fig:model1})
junction without ISOI.
The magnetic exchange field is oriented along $z$,
$\vec{h}=(0,0,h^z)$,  and therefore, solely the $z$ component
of spin current $J_x^{sz}(x,y)$ is nonvanishing. The junction length
and width are set to $d_F=2.0\xi_S$ and $W_F=2.0\xi_S$, respectively.
The
top panel exhibits the spin current variations along the $x$-position (the junction
length) at four differing locations along the junction width, $y=0.25\xi_S,
0.5\xi_S, 0.75\xi_S, 1.0\xi_S$. The bottom panel shows $J_x^{sz}(x,y)$
as a function of $y$-position along the junction width, at
$x=0.25\xi_S, 0.5\xi_S, 0.75\xi_S, 1.0\xi_S$. }
\end{figure}
%----------------------------------------------------------------------------
The Rashba SOI \cite{rashba_book_1}
can be described through spatial inversion asymmetries while the
Dresselhaus SOI \cite{dresselhaus_term} is described by bulk inversion
asymmetries in the crystal structure.\cite{winklwer_1,rashba_book_1}
Crystallographic inversion
asymmetries\cite{Ganichev} or lack of structural inversion
symmetries\cite{Miron_1,Garello_1,Duckheim1,Ganichev} in
heterostructures may cause the ISOIs considered here. For
example, strain can induce such inversion
asymmetries\cite{kato_1,cubic_rashba,Nakamura_1,Ganichev} and thus,
ISOIs, or the adjoining of two differing materials may generate the requisite
interfacial
SOIs\cite{Miron_1,Garello_1,Duckheim1,bergeret_so,Ganichev}.
Nonetheless, there is no straightforward method to measure SOIs
in a hybrid structure. One possible approach would be first principle
calculations\cite{Ast} in conjunction with spin transfer torque
experiments\cite{bergeret_so,Manchon,Ganichev}.
The intrinsic SOIs are often given by the first-order
 quasiparticle momentum, which is locked to their spins.
%------------------------------------------- figure 3 -------------------------
 \begin{figure*}[t!]
\includegraphics[width=17.5cm,height=8.10cm]{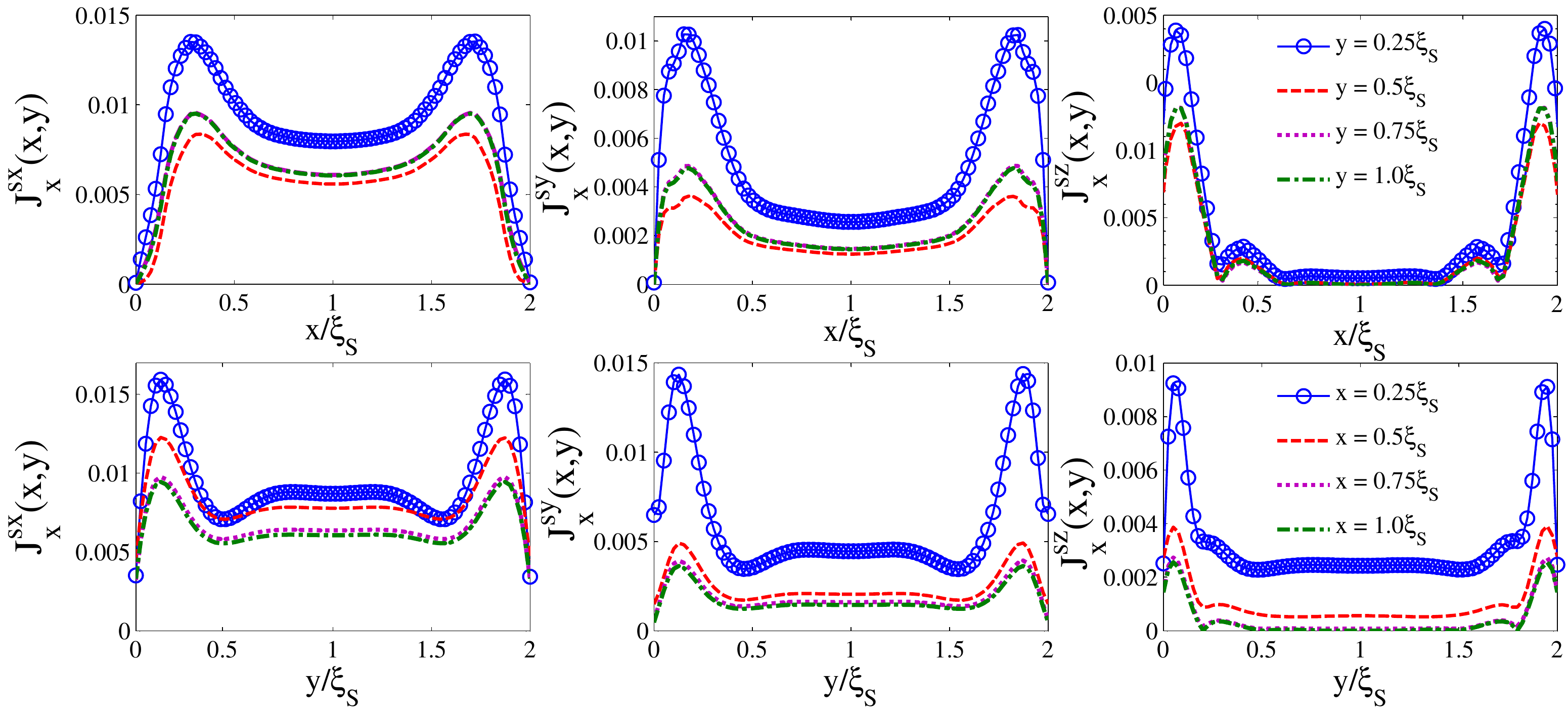}
\caption{\label{fig:hz_10_beta_5_alpha_0_xy} (Color online) Spatial
behavior of the three spin current components, $J^{sx}_x(x,y)$,
$J^{sy}_x(x,y)$, and $J^{sz}_x(x,y)$ in a uniformly magnetized Rashba \sfs
junction. The exchange
field of the ferromagnetic strip points along the $z$ direction:
$\vec{h}=(0,0,h^z)$ (see Fig.~\ref{fig:model1}).
The panels in the top row show the spin current
components, $J^{s\gamma}_x(x,y)$, as a function of $x$ at four
differing locations along the junction width: $y$=$0.25\xi_S$, $0.5\xi_S$,
$0.75\xi_S$, and $1.0\xi_S$. The bottom row exhibits $J^{s\gamma}_x(x,y)$
versus $y$ at $x$=$0.25\xi_S$, $0.5\xi_S$, $0.75\xi_S$, and $1.0\xi_S$.
}
\end{figure*}
%------------------------------------------------------------------------------
This linearized
approach is a simplification to the  more generic
picture dealing with higher orders of momentum,
\cite{winklwer_1,winkler_2,Minkov_1,Nakamura_1,chang_malsh_1,cubic_rashba}
which can be observed in e.g., engineered materials. \cite{cubic_rashba,Nakamura_1}
We here assume that ISOCs can be described by linear terms
in the carriers' momentum. \cite{rashba_book_1,dresselhaus_term}
Candidate materials to support spontaneous broken time-reversal and
broken inversion symmetries include electron liquids with ISOIs, which naturally
tend to have a Stoner-type magnetism at low densities, and a magnetically
doped topological insulator surface (or by directly coating
a topological insulator surface with magnetic insulators).\cite{cndid_1,cndid_2,cndid_3}
Other
promising candidates involve the ferromagnetic semiconductors (Ga,Mn)As, where
both the electronic structure and inherent magnetism make these materials well suited
for experimental
studies.\cite{cndid_5,major2,major1}
Our quasiclassical
approach allows us to study systems involving nontrivial magnetizations and
spin vector potentials with arbitrary spatial patterns.
We thus consider a finite-sized, uniformly
magnetized \f wire whose exchange field can take arbitrary
orientations. In order to determine systematically the behaviors of the spin and
charge currents, we consider three orthogonal magnetization
directions, namely along the $x$, $y$, and $z$ axes.
In addition, we
incorporate pure Rashba ($\alpha\neq 0, \beta=0$) and Dresselhaus
($\beta\neq0, \alpha= 0$) SOCs that allow isolation of their effects
relative to the physical quantities under study.
When finding solutions to
the Usadel equation, Eq.~(\ref{eq:usadel}), and  the
corresponding current densities [Eqs.~(\ref{eq:chargecurrentdensity}) and
(\ref{eq:spincurrentdensity})], we have added a small
imaginary part, $\delta\approx 0.01\Delta_0$, to the quasiparticles' energy,
$\varepsilon\rightarrow\varepsilon+i\delta$, to enhance  stability
of the numerical solutions. The imaginary part can be
physically viewed as accounting for inelastic scatterings.\cite{alidoust_1}
Due to the presence of the finite parameter $\delta$, we take the modulus of
the
currents in Eqs.~(\ref{eq:chargecurrentdensity}) and
(\ref{eq:spincurrentdensity}).
We normalize the quasiparticles' energy, $\varepsilon$, and exchange
field $\vec{h}$ by the gap, $\Delta_0$, at $T=0$. Also, all lengths
are measured in units of the superconducting coherence length $\xi_S$. In
our computations, we adopt natural units, so that $k_B=\hbar=1$.

To begin, we consider for comparison purposes,
an \sfs Josephson junction in the absence
of SOCs.
\cite{buzdin_rmp,bergeret1} The schematic of the
\sfs structure is depicted in Fig.~\ref{fig:model1}.
The parameters
$\zeta=4$, $|\vec{h}|=10\Delta_0$ and $d_F=2.0\xi_S$, ensure the
validity of low proximity limit considered throughout the paper. To
have absolute comparisons, we set $\vec{h}=(0,0,h^z)$ and compute
the charge and spin currents using Eqs.~(\ref{eq:chargecurrentdensity})
and (\ref{eq:spincurrentdensity}),
respectively. Figure \ref{fig:hz_10_beta_0_alpha_0_xy} exhibits the
spatial map of the spin current for $W_F=2.0\xi_S$
(see Fig. \ref{fig:model1}). Since the magnetization orientation is
fixed
along the $z$ direction, $J^{sz}_x(x,y=y_0)$ is the only
nonvanishing component of spin current for a given
fixed location $y_0$.
The top panel of Fig.~\ref{fig:hz_10_beta_0_alpha_0_xy}
illustrates the spatial variations
of $J^{sz}_x(x,y=y_0)$ along the junction length in the $x$ direction at
differing positions along the junction width: $y_0=0.25\xi_S$, $0.5\xi_S$,
$0.75\xi_S$, and $1.0\xi_S$. The macroscopic phase difference between the
\s electrodes is set at a representative value, i.e.,  $\varphi=\pi/2$.
The bottom panel in Fig.
\ref{fig:hz_10_beta_0_alpha_0_xy} shows $J^{sz}_x(x=x_0,y)$ as a
function of $y$, at $x_0=0.25\xi_S, 0.5\xi_S, 0.75\xi_S, 1.0\xi_S$. The
results demonstrate that the spin current is $y$ independent in such
hybrid junctions, namely $J^{sz}_x(x=x_0,y)=const.$ (we also have
found $J^{sz}_y(x,y)=0$). In other words, it is appropriate to view
this type of system
as an effectively one-dimensional junction.
The variation of
$J^{sz}_x(x,y)$ along the $x$ direction
is a consequence of
spin
torque transfer, and hence the spin current is
not a conserved
quantity.\cite{gorini_1,mishchenko_1,Nikolic,eschrig_1}
The spin current is maximal at the \fs interfaces and vanishes at
the middle of junction, $x=1.0\xi_S=d_F/2$. This is
contrast to
the charge supercurrent in the \f region, which is
conserved, and thus has a constant value within the entire \f strip (not shown).

%----------------------------------------- added analytical discussions ------------------
To identify some of the salient features
in  Fig.~\ref{fig:hz_10_beta_0_alpha_0_xy},
we consider now a simplified one-dimensional \sfs system,
which permits 
analytical expressions for the spin current density.
To this end, we  linearize the Usadel
equation, and incorporate the Kupriyanov boundary conditions, where the superconducting
electrodes have strong scattering impurities.
We also still assume that the magnetization is oriented along $z$:
$\vec{h}=(0,0,h^z)$.
Correspondingly, we define the dimensionless quantity, $\lambda_{\pm}=2i(\varepsilon\pm h^z)/\varepsilon_T$,
in which $\varepsilon_T$ is the Thouless energy, 
and the dimensionless $x$ coordinate, $\tilde{x}=x/d_F\in [0,1]$. After some straightforward calculations, we 
obtain the following expression for the charge current [Eq. (\ref{eq:chargecurrentdensity})]:
\begin{eqnarray}\label{eq:anl_cc}
&&\nonumber J_x^c(x,\varphi)=J_0^c \sin\varphi \int_{-\infty}^{+\infty} d\varepsilon\frac{2i\tanh(\varepsilon k_BT/2)}{\zeta^2\lambda_{+}\lambda_{-}}
\\&&\nonumber\bigg\{[s^*(-\varepsilon)]^2\Big( \lambda_{+}\csc\lambda_{-}+\lambda_{-}\csc\lambda_{+}\Big)+
[s^*(\varepsilon)]^2 \\&& \Big(\lambda_{+}\text{csch}\lambda_{-}+\lambda_{-}\text{csch}\lambda_{+} \Big)
\bigg\}.
\end{eqnarray}
The charge current in this case
is seen to exhibit
the usual $\sin\varphi$ odd-functionality in the
superconducting phase difference. 
Likewise, by substituting the solutions
into Eq.~(\ref{eq:spincurrentdensity}), we arrive at the
following expressions for the spin current components:
\begin{subequations}
\begin{eqnarray}\label{eq:jx}
&&J^{sx}_x(x,\varphi)\equiv 0,
\end{eqnarray}
\begin{eqnarray}\label{eq:jy}
&&J^{sy}_x(x,\varphi)\equiv 0,
\end{eqnarray}
\begin{eqnarray}\label{eq:jz}
    &&\nonumber J^{sz}_x(x,\varphi)=
    J_0^s\int_{-\infty}^{+\infty} d\varepsilon\frac{2\tanh(\varepsilon k_BT/2)}{\zeta^2\lambda_{+}\lambda_{-}}
    \\&&\nonumber\Big\{ [s^*(\varepsilon)]^2\lambda_{+}\cosh
    2\tilde{x}\lambda_{-}\text{csch}\lambda_{-}\cos\varphi+[s^*(-\varepsilon)]^2\\ && \nonumber
     \Big(\lambda_{+}\csc^2\lambda_{-}(\cos\lambda_{-}+\cos\varphi)\sin[\lambda_{-}(1-2\tilde{x})]
    -\\&&\nonumber \lambda_{-}\csc^2\lambda_{+}(\cos\lambda_{+}+\cos\varphi)\sin[\lambda_{+}(1-2\tilde{x})]
    \Big)+\\&&\nonumber [s^*(\varepsilon)]^2\Big(
     \lambda_{+}\coth\lambda_{-}\text{csch}\lambda_{-}(\sinh[\lambda_{-}(1-2\tilde{x})]-
    \\&&\nonumber \sinh 2\tilde{x}\lambda_{-}\cos\varphi)-\lambda_{-}\text{csch}^2\lambda_{+}(\cosh\lambda_{+}+\cos\varphi)\\&&\sinh[\lambda_{+}(1-2\tilde{x})]
    \Big\}.
\end{eqnarray}
\end{subequations}
Equations (\ref{eq:jx})-(\ref{eq:jz}) clearly demonstrate
that the only nonvanishing component of spin current is $J^{sz}_x$,
which is consistent with
the exchange field aligned along $z$.\cite{ma_kh_soc_sa}
From Eq.~(\ref{eq:jz}), it is also evident that $J^{sz}_x$
is an odd function of the coordinate $\tilde{x}$ relative to the middle of the junction
(and thus  vanishes there), and an even function of
the phase difference,  $\varphi$.
These features are entirely  consistent with the numerical results
seen in Fig.~\ref{fig:hz_10_beta_0_alpha_0_xy}.

%-----------------------------------------------------------------------

%------------------------------------------- figure 4 -------------------------
\begin{figure*}[t!]
\includegraphics[width=17.5cm,height=8.10cm]{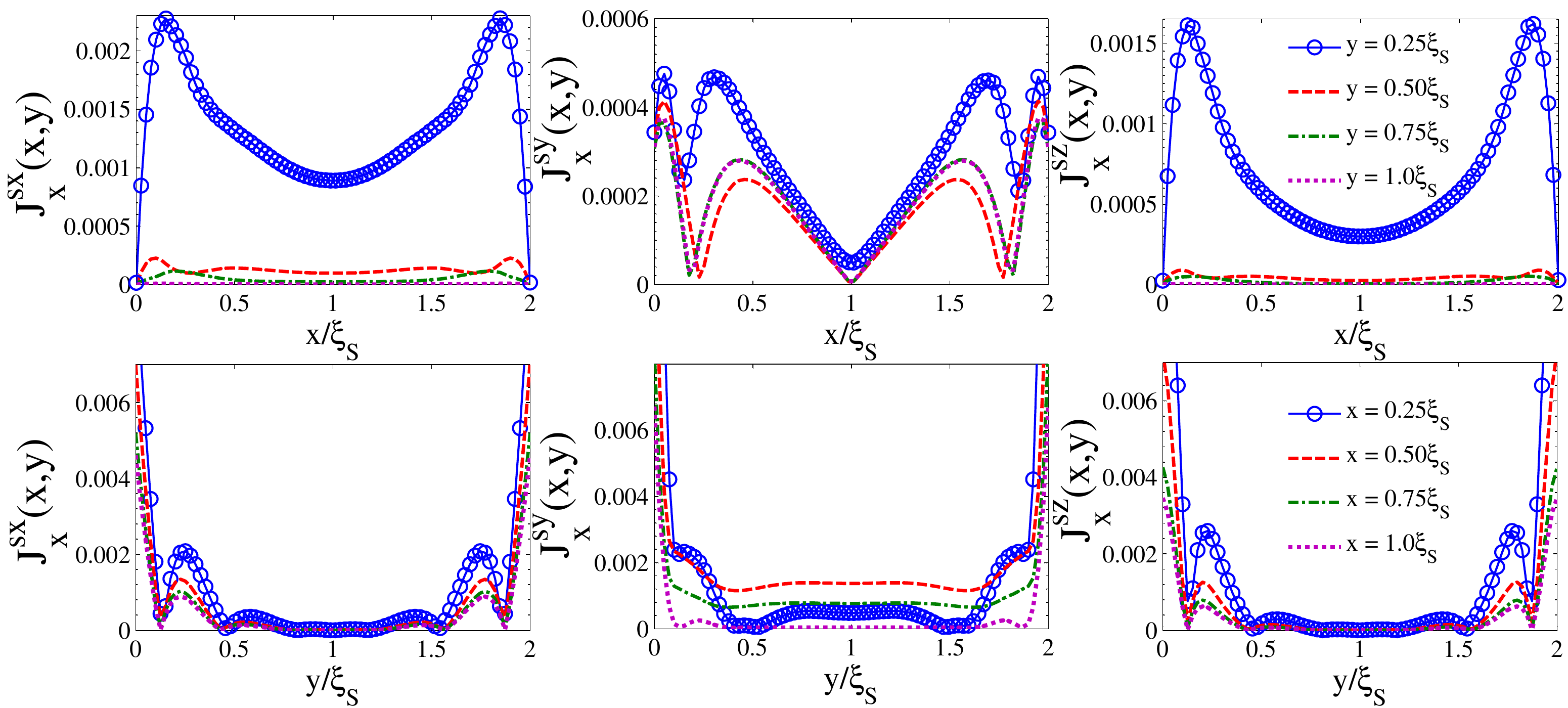}
\caption{\label{fig:hy_10_beta_5_alpha_0_xy} (Color online) Spatial
profiles of the spin current components; $J^{sx}_x(x,y)$,
$J^{sy}_x(x,y)$, and $J^{sz}_x(x,y)$ in an \sfs system. The
Rashba ferromagnetic wire's width and length are equal
to $W_F=d_F=2.0\xi_S$. The exchange field of the
ferromagnetic strip is
fixed along the $y$ direction, $\vec{h}=(0,h^y,0)$. \textit{Top row}: spatial behavior of
$J^{s\gamma}_x(x,y)$ along the junction length, $x$, at
$y=0.25\xi_S$, $0.5\xi_S$, $0.75\xi_S$, $1.0\xi_S$. \textit{Bottom
row}: spatial variations of $J^{s\gamma}_x(x,y)$ along the junction
width in the $y$ direction at $x=0.25\xi_S$, $0.5\xi_S$,
$0.75\xi_S$, $1.0\xi_S$. }
\end{figure*}
%------------------------------------------------------------------------------
We now incorporate Rashba and Dresselhaus SOCs,
while keeping the magnetization orientation intact along the
$z$ direction.
The ISOIs are confined within the \f region and are not present
within the \s electrodes. Through exhaustive numerical  investigations,
we have found several symmetries among the components
of spin current (discussed below) at three particular directions of the exchange field.
Due
to the symmetries available among the spin current components, we
focus here on Rashba SOC. We emphasize that similar conclusions can be drawn for
 Dresselhaus SOC
through the symmetries described below.
Figure
\ref{fig:hz_10_beta_5_alpha_0_xy} exhibits the spatial profiles for
the spin current density components, $J^{sx}_x(x,y)$, $J^{sy}_x(x,y)$, and
$J^{sz}_x(x,y)$. A square ferromagnetic strip is considered, with
$d_F=W_F=2.0\xi_S$, and the superconducting phase difference
is equal to $\varphi=\pi/2$.
The Rashba SOC coefficient is
set to a representative value $\alpha=2.0\xi_S$, without loss of
generality.\cite{bergeret_so}
The top set of panels show
$J^{s\gamma}_x(x,y=y_0)$ [$\gamma=x,y,z$] as a function of the $x$ coordinate at
$y_0=0.25\xi_S$, $0.5\xi_S$, $0.75\xi_S$, and $1.0\xi_S$.
Whereas the bottom
panels represent the same quantities, but now as a function of $y$ at
$x_0=0.25\xi_S$, $0.5\xi_S$, $0.75\xi_S$, and $1.0\xi_S$.
As seen in Fig.~\ref{fig:model1}, the junction length and width are parallel to the
$x$ and $y$ axes, respectively. The components $J^{sx}_x(x,y=y_0)$ and
$J^{sy}_x(x,y=y_0)$, shown in the top row of Fig.~\ref{fig:hz_10_beta_5_alpha_0_xy},
demonstrate that these spin current densities vanish at the \fs contacts. This finding is
consistent with previous works involving
nonsuperconducting
heterojunctions
\cite{gorini_1,mishchenko_1,Nikolic,kato_1,malsh_sns}.
The $z$ component, $J^{sz}_x(x,y=y_0)$,
however exhibits
opposite behavior, and is nonzero at the \fs
contacts due to the
exchange field, which is oriented along the $z$ axis.
Similarly, as seen in Fig.~\ref{fig:hz_10_beta_0_alpha_0_xy},
$J^{sz}_x(x,y=y_0)$ is finite at
the \fs interfaces near the \s reservoirs.
One of the most important
features of the results is seen
in the top panels of Fig.~\ref{fig:hz_10_beta_5_alpha_0_xy},
where  two peaks in
$J^{s\gamma}_x(x,y)$ emerge near the \fs contacts.
We restrict the
spatial profiles to $0<x<d_F/2$ and $0<y<W_F/2$,
since the results are symmetrical with respect to
$x=1.0\xi_S=d_F/2$ and $y=1.0\xi_S=W_F/2$,
so that the maxima of $J^{s\gamma}_x(x,y=y_0)$
occurs near the
edges of the \f wire [at $x=0$, and $x=d_F$].
Turning to the bottom row of panels in  Fig.~\ref{fig:hz_10_beta_5_alpha_0_xy}, we see that
$J^{s\gamma}_x(x=x_0,y)$ are nonzero at the vacuum boundaries, $y=0$,
and $y=W_F$. Here also the largest values in the spin current density
components take place near the transverse edges of the \f wire ($y=0$, and $y=W_F$).
The magnitude of the spin current densities at $x=0.25\xi_S$ are
generally larger than the other $x$ positions, in agreement with the
results of $J^{s\gamma}_x(x,y=y_0)$ shown in the top row of panels.

We now consider the effects of changing the magnetization alignment
in the ferromagnet. Thus,
Fig.~\ref{fig:hy_10_beta_5_alpha_0_xy} represents the same Rashba
spin-orbit coupled \sfs junction as in Fig.~\ref{fig:hz_10_beta_5_alpha_0_xy},
except  the magnetization of the \f
wire is now oriented along the $y$ axis.
This specific direction of $\vec{h}$ leads to $J^{sx}_x(x,y=y_0)=J^{sz}_x(x,y=y_0)=0$ at the
\fs interfaces and the spin current densities peak near the edges of \f wire.
The spin current density
$J^{sy}_x(x,y=y_0)$ however is nonzero at the \fs contacts
similarly to $J^{sz}_x(x,y=y_0)$ in the configuration where the
magnetization points along the $z$ direction (Fig.~\ref{fig:hz_10_beta_5_alpha_0_xy}).
As mentioned earlier, this
nonvanishing behavior is directly related to
the exchange field direction
which lies now parallel to the $y$ axis.
Examining $J^{s\gamma}_x(x=x_0,y)$ in the bottom row panels in
Fig.~\ref{fig:hy_10_beta_5_alpha_0_xy}, the maximal values of
$J^{s\gamma}_x(x=x_0,y)$ take place near the vacuum boundaries, i.e. $y=0$,
and $y=W_F$. Our investigations demonstrate similar qualitative trends for
the components of $J^{s\gamma}_x(x=x_0,y)$ when the magnetization
resides along the $x$ axis.
Note
that the transverse components of the spin currents are nonzero
inside the ISO coupled ferromagnetic wire, i.e.,
$J^{s\gamma}_y(x,y)\neq 0$, and vanish at the
vacuum boundaries ($y=0, W_F$).
We mainly focus here on the $J^{s\gamma}_x(x,y)$,
since the longitudinal components contain the relevant information
needed to describe and understand the accumulation of spin
current densities at the edges of the structures.

\begin{figure}[b!]
\includegraphics[width=6.0cm,height=2.0cm]{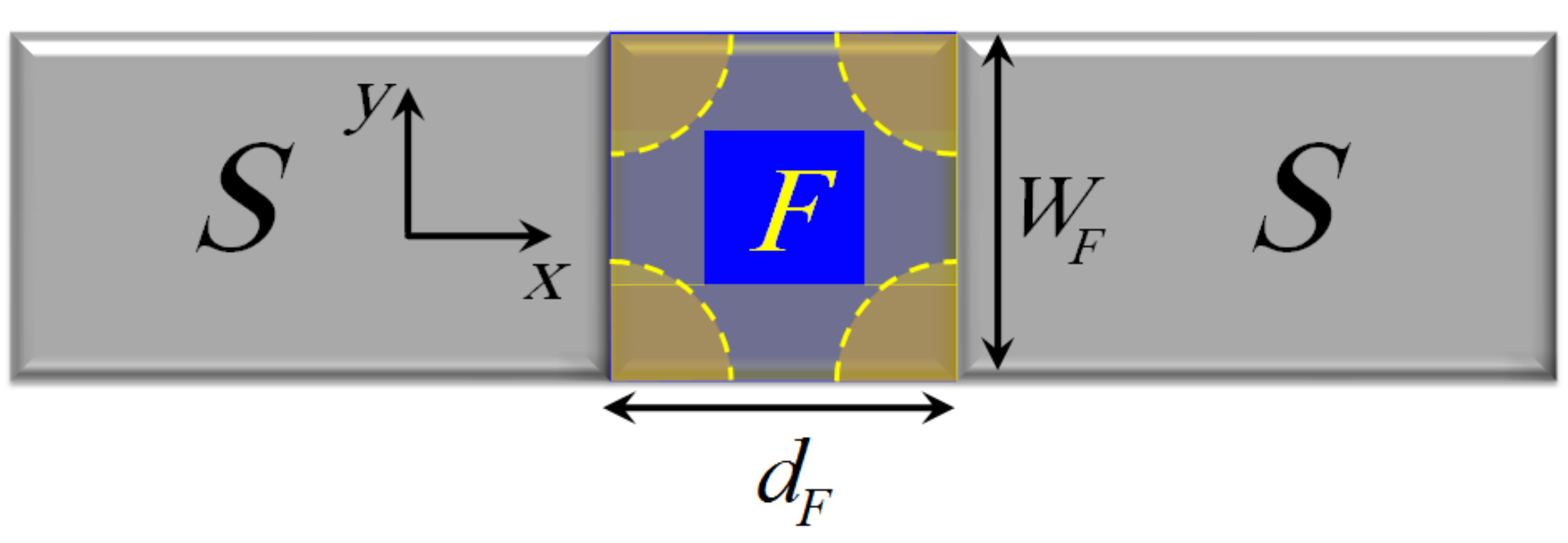}
\caption{\label{fig:model2} (Color online) Qualitative illustration
of the edge spin current densities in a Rashba or
Dresselhaus spin-orbit coupled \sfs junction.
The light yellow ribbons display edge regions with maximum spin current densities.
The induced spin currents can be
considered as a response of the intrinsic spin-orbit coupled system to the
presence of an exchange field (the combination of spontaneously broken
time-reversal symmetries and the lack of inversion symmetries). As shown in Fig.~\ref{fig:model1}, the
exchange field of the ferromagnetic wire is uniform and can take
arbitrary directions. The regions that carry maximal accumulation of
spin currents are qualitatively shown by
the semicircular regions.}
\end{figure}

Considering now
the previous characterization of the spin current
components
in systems with ISOCs,
we schematically summarize the spatial
maps in Fig.~\ref{fig:model2}
for $\vec{J}^{s\gamma}(x,y)$.
The
largest amplitudes of $\vec{J}^{s\gamma}(x,y)$  reside
near the edges of the  \f strip, i.e. $x=0, d_F$ and $y=0, W_F$. We have
qualitatively marked these regions by light yellow ``ribbons". Therefore,
the overlap of maximal amplitudes take place near the corner regions
of the \f strip. We have marked these areas by dashed curves and with a deeper
yellow color. The spatial profiles found here are
qualitatively similar to the existence of edge spin currents
found in  nonsuperconducting heterojunctions with ISOIs,
\cite{Murakami_1,hirsh_1,kato_1,Sinova_1,mishchenko_1,chazalviel,
Nikolic,Onoda}
except with one crucial difference:  here spin accumulation
at the edges arises in the absence of an external field.
As mentioned in other works\cite{malsh_sns}, the spin accumulation is a
signature of the spin Hall effect. Therefore, the predicted spin
accumulation in this paper might be measurable through optical
experiments\cite{kato_1}, such as through Kerr rotation microscopy\cite{kato_1},
where spatial profiles of
the spin
polarizations near the edges can be imaged.
An alternate experimental proposal involves
multiterminal devices \cite{Nikolic,mishchenko_1}.
When
transverse leads are attached to the lateral edges of a
two-dimensional \sfs junction (borders at $y=0$, $y=W_N$ in Fig.
\ref{fig:model1}(a)), the spin accumulations at the $F$ wire's edges
inject spin currents into the leads.
\cite{Nikolic,mishchenko_1}
The transversely injected
spin currents into the lateral leads in turn may induce
a voltage drop between the
additional leads.\cite{Nikolic,mishchenko_1}

%------------------------------------------- figure 5 -------------------------
%--------------------------------------------------------------------------

%------------------------------------------- figure 6 -------------------------
\begin{figure*}[t!]
\includegraphics[width=17cm,height=8.30cm]{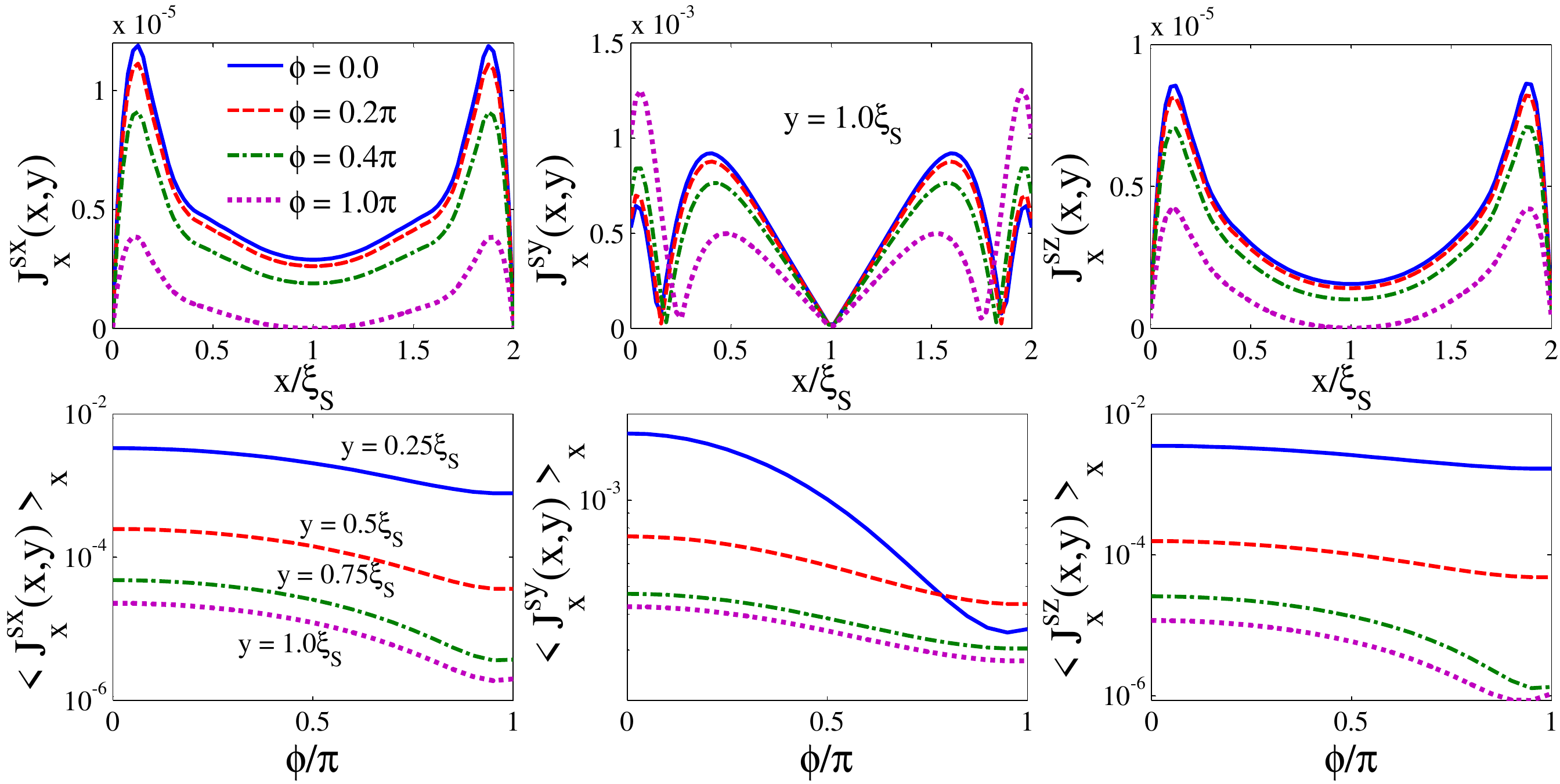}
\caption{\label{fig:hy_10_beta_5_alpha_0_x_phi} (Color online) Spin
current components for  differing values of the superconducting phase
difference $\varphi$  in a
Rashba spin-orbit coupled \sfs Josephson junction.
The top row of panels shows the spatial variations of
$J_x^{s\gamma}(x,y)$ as a function of $x$ along the junction length
at $\varphi=0, 0.2\pi, 0.4\pi$, and $1.0\pi$. The location along the
junction width is fixed at the middle of the junction, $y=W_F/2=1.0\xi_S$.
The bottom row of panels represents the spatially averaged spin current
components over the junction length (denoted by
$\langle J_x^{s\gamma}(x,y)\rangle_x$) vs $\varphi$.
The average is performed along four positions: $y$=$0.25\xi_S$,
$0.5\xi_S$, $0.75\xi_S$,  and $1.0\xi_S$. The ferromagnetic wire is a
square strip with $d_F=W_F=2.0\xi_S$, and exchange field $\vec{h}=(0,h^y,0)$. }
\end{figure*}
%--------------------------------------------------------------------------

The Josephson effect is a significant example of  a macroscopic quantum
phenomenon and is of fundamental importance in determining the
properties of  dissipationless coherent transport.
Thus, the behavior of
the spin currents upon varying the macroscopic phase
difference is crucial to experiments and applications utilizing  spin-Hall
effects and spin transport. In Fig.~\ref{fig:hy_10_beta_5_alpha_0_x_phi},
we therefore study the spin
current components as a function of the macroscopic phase
difference, $\varphi$, between the \s banks. We consider the
parameter set used in Fig.~\ref{fig:hy_10_beta_5_alpha_0_xy},
including
$\vec{h}=(0,h^y,0)$. In the top set of panels,
the spatial variations of
$J^{s\gamma}_x(x,y)$ are plotted at $\varphi=0$, $0.2\pi$, $0.4\pi$, and $1.0\pi$.
We have also chosen
a representative position along the junction width, corresponding to
$y=1.0\xi_S=W_F/2$,  which simplifies the analysis while
maintaining the generality of the discussion.
Although $J^{s\gamma}_x(x,y)$
has a  minimum at $y=1.0\xi_S$, it exhibits the same trends
as a function of
$\varphi$ compared to the other positions inside the \f wire.
By increasing
the superconducting phase difference from $0$ to $\pi$, the
amplitudes of the spin current components decrease overall.
In the bottom
set of panels of Fig.~\ref{fig:hy_10_beta_5_alpha_0_xy}, we illustrate
$\langle J^{s\gamma}_x(x,y)\rangle_x$ as a function of $\varphi$, at $y=0.25\xi_S$, $0.5\xi_S$,
$0.75\xi_S$, and $1.0\xi_S$. Here, we denote the spatial average over the $x$
coordinate from $0$ to $d_F$ by $\langle...\rangle_x$.
In order to better
visualize the averaged profiles, we use logarithmic scales in
the vertical axes of the bottom row of panels.
As seen, the three
components of spin current $\langle J^{s\gamma}_x(x,y)\rangle_x$ are
even-functions of $\varphi$, with a period of $2\pi$, namely
$J^{s\gamma}(2n\pi+\varphi)=J^{s\gamma}(-\varphi)$, $n\in
\mathbb{Z}$. This is contrary to the charge supercurrent
which is an odd-function of $\varphi$, i.e.,
$J^{c}(2n\pi+\varphi)=-J^{c}(-\varphi)$
regardless of a finite phase-shift $\varphi_0$
\cite{Konschelle,nazar_1,buzdin_phi0}. These
findings are entirely consistent with previous studies of  \sfs
Josephson junctions with inhomogeneous magnetization patterns\cite{alidoust_1}.
We here remark that an additional phase-shift $\varphi_0$ may appear
in such junctions due to the coupling of exchange field and ISOIs.
\cite{Konschelle,nazar_1,buzdin_phi0}
Nonetheless, the explicit current-phase
relations simply undergo a shift in $\varphi_0.$\cite{nazar_1,buzdin_phi0}
According to the current-phase relations,
the charge supercurrent vanishes at certain $\varphi$ that
is quite different than the behavior of the spin current components
which clearly show nonzero values at the same $\varphi$. Therefore, these
differences in charge
and spin currents allows for an examination
of edge spin currents without any
net charge current in an ISO coupled \f wire sandwiched between two \s banks.

We are now in a position to discuss  symmetries
that may arise among the spin current density components for
differing magnetization orientations in systems with
either Rashba or
Dresselhaus SOCs.
Our investigations have found that the out-of-plane spin current, $\vec{J}^{sz}(x,y)$,
remains unchanged upon exchanging the Rashba and Dresselhaus SOCs,
regardless of the magnetization orientation. This follows from the
form of the spin vector potential discussed at the beginning of this section.
However, this picture
changes for the in-plane $\vec{J}^{s\{x,y\}}(x,y)$ components.
The $x$ and $y$ components of the spin current become interchanged
when transforming from one type of spin-orbit interaction to another.
Precisely speaking, by going from Rashba to Dresselhaus spin-orbit coupling,
one simply needs to exchange indices $x$ and $y$ in the components of
both the exchange field and the spin current. Otherwise, everything stays the same.
By making use of the simple transformation rules described, one
can easily deduce the results of Dresselhaus spin-orbit coupled
systems from the plots presented for Rashba spin-orbit coupled \sfs systems
shown
in Figs. \ref{fig:hz_10_beta_5_alpha_0_xy}, \ref{fig:hy_10_beta_5_alpha_0_xy}, and
\ref{fig:hy_10_beta_5_alpha_0_x_phi}.

To conclude this section, we briefly
discuss the importance of having a magnetic element in the Josephson junction
for the effect of spin current edge accumulation  to take place spontaneously.
We thus take
the limiting case of $\vec{h}=0$ in our previous
calculations above involving \sfs junctions.
Using otherwise the same geometrical and material parameters,
this case was found to
produce no spin current, $\vec{J}^{s\gamma}(x,y)=0$,
in the presence
of Rashba ($\alpha\neq 0, \beta=0$) and/or Dresselhaus ($\beta\neq
0, \alpha=0$) SOIs. These findings are
consistent with previous works, \cite{malsh_sns} where several
simplifying approximations were employed for
Rashba-based \sns systems. Examining also the charge
supercurrent, $\vec{J}^{c}(x,y)$, for both the Rashba and Dresselhaus interactions,
we observed
a uniform
spatial map for the charge current density for all
$\varphi$,  with  $J^{c}_x=const.$, and $J^{c}_y=0$.
In other words, the spin-dependent fields cannot
induce transverse charge supercurrents in a diffusive \sns
junction. This is in stark contrast to its ballistic \sns counterpart,
where a transverse charge supercurrent (that is, equivalent to a
supercurrent flowing along the $y$ direction in our configuration
depicted in Fig. \ref{fig:model1}) was theoretically predicted due to the presence
of intrinsic SOIs\cite{yokoyama_1}.

\section{Conclusions}\label{sec:conclusion}

We have theoretically studied the behavior of spin and charge currents in a
finite-size two-dimensional \sfs Josephson junction with intrinsic spin-orbit couplings.
We  utilized a two-dimensional
Keldysh-Usadel quasiclassical approach that incorporates
a generic spin-dependent vector potential. Our results
demonstrate that the combination of a uniform magnetization and
ISOIs drives the spin currents
which spontaneously accumulate at the
\f wire's edges. The corners of the $F$ wire were shown to
host the maximum density of
spin currents. (As demonstrated in Ref. \onlinecite{ma_kh_soc_sa},
similar edge phenomena can be found in finite-size two-dimensional intrinsically spin orbit coupled
\sns junctions with a single spin active interface. Additionally, it
was shown that maximum singlet-triplet conversions take place at the
corners of N wire nearest the spin active interfaces\cite{ma_kh_soc_sa}.)
Our
investigations
show that the spontaneous edge accumulation of the spin currents  are robust
and can exist at all magnetization orientations, independent of the actual type of ISOIs.
Our investigations
have also
found several symmetries among the spin current
components upon varying magnetization orientations coupled to a
Rashba or Dresselhaus SOI. By varying the superconducting phase difference, $\varphi$,
between the \s banks, we determined
the spin
and charge currents as a function of phase difference. We have found that
net spin currents therefore emerge
and accumulate spontaneously at the edges, in the absence
of charge
flow, when properly modulating $\varphi$ in finite-size two-dimensional intrinsically
spin-orbit coupled \sfs hybrid structures.
This work can be viewed as  complementary to  previous
studies involving  edge spin currents in
non-superconducting spin-orbit coupled structures where
externally imposed fields were required
\cite{gorini_1,malsh_severin,malsh_sns,hirsh_1,kato_1,mishchenko_1,Nikolic}.
We have shown that remarkably,
edge spin currents can be spontaneously driven by the coupling of intrinsic
properties of a system, i.e. spontaneously broken time-reversal and the lack of inversion symmetries
in the {\it absence} of any externally imposed  field.

\acknowledgments
We would like to thank G. Sewell for helpful discussions in the
numerical parts of this work. We also thank F.S. Bergeret for
valuable comments, suggestions, and numerous discussions
which helped us to improve the manuscript.
K.H. is supported in part by ONR and by a grant of supercomputer resources
provided by the DOD HPCMP.

\appendix

\section{Pauli Matrices}\label{app:pauli}
In Sec.~\ref{sec:theor} we introduced the Pauli matrices in the spin
space and denoted them by $\vec{\sigma}=\big(\sigma^x, \sigma^y,
\sigma^z\big)$, $\vec{\tau}=\big(\tau^x, \tau^y, \tau^z\big)$, and
$\vec{\nu}=\big(\nu^x, \nu^y, \nu^z\big)$.
\begin{align}
&\sigma^x = \begin{pmatrix}
0 & 1\\
1 & 0\\
\end{pmatrix},\;
\sigma^y = \begin{pmatrix}
0 & -i\\
i & 0\\
\end{pmatrix},\;
\sigma^z = \begin{pmatrix}
1& 0\\
0& -1\\
\end{pmatrix},\;
\sigma^0 = \begin{pmatrix}
1 & 0\\
0 & 1\\
\end{pmatrix}.\nonumber
\end{align}
We also introduced the $4\times 4$ matrices
$\vec{\hat{\rho}}=(\hat{\rho}_1, \hat{\rho}_2, \hat{\rho}_3)$:
\begin{align}
\nonumber &\hat{\rho}_1 =
\begin{pmatrix}
0 & \sigma^x\\
\sigma^x & 0 \\
\end{pmatrix},\;
\hat{\rho}_2 =  \begin{pmatrix}
0 & -i\sigma^x\\
i\sigma^x & 0 \\
\end{pmatrix},\;
\hat{\rho}_3 = \begin{pmatrix}
\sigma^0 & 0\\
0 & -\sigma^0  \\
\end{pmatrix}.
\end{align}
Following Ref. \onlinecite{alidoust_1}, we define $\tau^\gamma$,
$\nu^\gamma$, and $\hat{\rho}_0$ as follows;
\begin{align}
 \tau^\gamma = \begin{pmatrix}
\sigma^\gamma & 0\\
0 & \sigma^\gamma\\
\end{pmatrix},\; \nu^\gamma = \begin{pmatrix}
\sigma^\gamma & 0\\
0 & \sigma^{\gamma\ast}\\
\end{pmatrix},\;\hat{\rho}_0 = \begin{pmatrix}
\sigma^0 & 0\\
0 & \sigma^0 \\
\end{pmatrix},\nonumber
\end{align}
to unify our notation throughout the paper $\gamma$ stands for
$x,y,z$.

\end{document}